\documentclass[
  reprint,
  showpacs,
  showkeys,
  amsmath,
  amssymb,
  aps,
  prmaterials
]{revtex4-2}
\usepackage{graphicx} 
\usepackage{rotating}
\usepackage{color}
\usepackage{float}
\usepackage{dcolumn}  
\DeclareGraphicsExtensions{.pdf,.png,.jpg}

\begin{document}

\preprint{Yamashita et al., Mn$_2$VZ (Z = Al, Ga)}

\title{First-principles study of magnetic and spin-dependent transport properties of  
       Mn$_2$VZ (Z = Al, Ga) with negative spin polarization using 
			 { {a}} disordered local moment approach at finite temperatures}

\author{Shogo Yamashita}
\email{shogo.yamashita@cpfs.mpg.de}
\affiliation{Max-Planck-Institute for Chemical Physics of Solids,
             D-01187 Dresden, Germany}
						
\author{Esita Pandey}
\affiliation{Max-Planck-Institute for Chemical Physics of Solids,
             D-01187 Dresden, Germany}

\author{Gerhard H. Fecher}
\affiliation{Max-Planck-Institute for Chemical Physics of Solids,
             D-01187 Dresden, Germany}

\author{Claudia Felser}
\affiliation{Max-Planck-Institute for Chemical Physics of Solids,
             D-01187 Dresden, Germany}

	
\author{Atsufumi Hirohata}
\affiliation{Tohoku University,
Center for Science and Innovation in Spintronics,
Sendai 980-8577, Japan}
\affiliation{Max-Planck-Institute for Chemical Physics of Solids,
D-01187 Dresden, Germany}	
						

\begin{abstract}

First-principles studies were performed on two Mn-based  ferrimagnetic Heusler
compounds with $L2_1$ and $B2$ structures, that { is,} Mn$_2$VZ (Z = Al or Ga). The
aim was to investigate their magnetic properties, electronic structures, and
spin-resolved longitudinal conductivity at finite temperatures. Density
functional theory (DFT) and functional integral theory were { used}. This
approach incorporates transverse spin fluctuations through { a} disordered local
moment method and the coherent potential approximation.
In all cases, the calculated theoretical Curie temperatures { were} lower than the experimental values. 
Alloys with a $B2$ structures { exhibit} higher Curie temperatures { compared to} compounds with an $L2_1$ structures. Calculations of the temperature
dependence of the density of states (DOS) indicate that the half-metallic
electronic structure collapses { owing} to the renormalization of transverse spin fluctuations at a finite temperatures. 
However, { the} spin-resolved longitudinal conductivities demonstrated an improved spin polarization, particularly for Mn$_2$VGa with an $L2_1$ structure.
This { result} contradicts predictions based on the temperature-dependent DOS. 
The competition between { the} metallic transitions, which are caused by { a modification} of the DOS, and scattering coming from spin-disorder explains this phenomenon. 
Both of these { effects} are induced by transverse spin fluctuations. 
Additionally, the results show that half-metallicity, as defined by { the} DOS or conductivity, is inconsistent at finite temperatures.
Finally, the total energy landscape of  the paramagnetic state was calculated using the fixed spin moment method to investigate the strength of { the} longitudinal spin fluctuations. 
These results { suggest} that the alloys may exhibit strong longitudinal spin fluctuations.

\end{abstract}


\maketitle

\section{Introduction}

Half-metallic Heusler alloys, such as Co$_{2}$-based Heusler alloys~\cite{Kubler1983,Galanakis2002,Galanakis2006}, are essential materials
in the field of spintronics, { owing to} their unique electronic structure. One spin 
channel is metallic, { whereas} the other spin channel is semiconducting. 
Several Mn-based Heusler alloys exhibit { interesting} properties. 
For example, Mn$_2$CoAl~\cite{Ouardi2013} is a spin-gapless semiconductor and
Mn$_{1.5}$V$_{0.5}$FeAl~\cite{Stinshoff2017} is a fully compensated ferrimagnet that exhibits no a net magnetization while maintaining a half-metallic electronic structure.

Among the Heusler compounds it is known that Mn$_2$VAl and Mn$_2$VGa are known
{as} half-metallic ferrimagnets that exhibit a negative spin polarization.
Their Fermi levels { are} located in the band gap formed in their majority spin
channel~\cite{Ishida1984,Weht1999,Ozdogan2006,Wollmann2014,Fan2020}. 
According to the electron-counting rule based on the localized part of the Slater--Pauling curves, the number of valence electrons ($n_v$) in the primitive cell of
Mn$_2$VAl and Mn$_2$VGa is 22. 
This {property} results in an expected total magnetic moment of 2~$\mu_B$~\cite{Wollmann2004}. 
This value is lower than that of Co$_{2}$-based half-metallic ferromagnetic Heusler compounds, which have { more}  valence electrons. 
Examples include Co$_2$MnSi and Co$_2$FeSi with $n_v=29$ and 30, respectively. 

{ Because of their} low magnetic moments, stray fields are expected to be suppressed.
Reducing the critical current should improve magnetization switching with spin-transfer torque and minimize electronic power consumption. 
Recently, current-perpendicular-to-plane giant magnetoresistance (CPP-GMR) devices were fabricated using Mn$_2$VAl and Mn$_2$VGa. 
These devices, exhibit large negative magnetoresistance of $-4.4$\% for Mn$_2$VAl~\cite{Sato2024_2} and $-1.8$\% for Mn$_2$VGa~\cite{Sato2024} at room temperature, respectively. 
A negative spin polarization { was} confirmed for these alloys.

From a spintronics perspective, a theoretical understanding of the temperature dependence of material-specific properties is highly desirable. 
These properties include spin polarization and the influence of chemical disorder on the electronic structure and related transport properties. 
This knowledge will provide a fundamental understanding of Heusler compounds and alloys, { and} assist in the fabrication of superior spintronics devices.

Theoretical knowledge based on { density functional theory} (DFT) is mostly limited to zero temperature.
Therefore, attempts have been made to extend { this method to include} finite temperature { effects}. 
To study the { magnetic} properties of Heusler alloys at finite temperatures,
Heisenberg-type spin Hamiltonians or their extensions are typically used, as in references~\cite{Sasioglu2004, Rusz2006}. 
However, as several { researches} have { indicated}, $d$ electrons in metallic systems can exhibit both itinerant and localized properties. 
For Mn$_2$VAl, x-ray spectroscopy confirmed the  itinerant nature of $d$ electrons at the Mn atoms and the localized nature of those at the V atoms~\cite{Nagai2018,Umetsu2019}. 
A similar situation { is} expected { in} the isostructural Mn$_2$VGa.
Therefore, a theoretical approach that incorporates both { of these} properties is necessary to describe metallic magnetic materials, including Mn-based Heusler alloys, at finite temperatures. 
Based on this concept, theoretical schemes { for treating} spin fluctuations in metallic magnets have been developed using the Hubbard model (see references~\cite{Cyrot,Hubbard1,Hubbard2,Hasegawa1,Hasegawa2,Hasegawa3}). 
These schemes have been translated into first-principles methods (see references~\cite{Oguchi,Pindor,Staunton3,Gyorrfy,StauntonPRL1992}).

The aim of this study {was} to examine the temperature-dependent magnetic properties, electronic structures, and spin-resolved longitudinal conductivities of Mn$_2$VAl and Mn$_2$VGa. 
The finite-temperature properties were calculated using functional integral theory based on DFT. 
This theory expresses {the} transverse spin fluctuations at finite temperatures and incorporates the itinerant and localized properties of $d$ electrons. 
The disordered local moment (DLM) method was employed  to address transverse spin fluctuations at finite temperatures (see references~\cite{Oguchi,Pindor,Staunton3,Gyorrfy,StauntonPRL1992}). 
The spin-disordered electronic structure of the systems at finite temperatures {was} evaluated using coherent potential approximation (CPA). 
 { Using}  this scheme,  {We can} calculate the temperature dependence of  {the} electronic structure and transport properties from  { the} electronic structures based on the DFT, and the linear response theory. 
Accordingly, the temperature dependence of  {the} magnetization, DOS, and spin-resolved longitudinal conductivities of Mn$_2$VAl and Mn$_2$VGa  { was} calculated for both  { the} $L2_1$ and $B2$ structures. 
Finally, the strength of  { the} longitudinal spin fluctuations  {was explored}  by investigating the total energy landscape in paramagnetic states. 
The aim {was} to briefly study the  {properties} of spin fluctuations  {in}  both alloys.
 { The findings of this study clarify how transverse spin fluctuations alter the electronic structure and affect spin-resolved conductivity at finite temperatures. } 
\section{Details of the calculations}

First-principles calculations based on DFT were used to investigate the temperature dependence of {the} magnetization, electronic structure, and transport properties. 
The electronic structure at zero temperature was calculated using a computer program based on the Korringa--Kohn--Rostoker (KKR) method combined with the atomic sphere approximation (ASA)~\cite{Abrikosov1993,Ruban1999}. 
The potential parameters of the tight-binding linear muffin-tin orbital (TB-LMTO) method were extracted from the calculation results. After obtaining  the TB-LMTO potential parameters for Mn$_2$VAl and Mn$_2$VGa, functional integral theory based on the TB-LMTO method was applied~\cite{Sakuma2000,Andersen,Turek,Kudrnovsky,Skriver}.
By using functional integral theory, the probability of spin transverse fluctuations $\omega(T,{\bf e}_i)$ at temperature $T$ and site $i$ within the single-site approximation, and from the electronic structure obtained with DFT. 
This theory incorporates transverse spin fluctuations at finite temperatures {by} using a vector-type DLM picture. {The evaluation of} the functional integrals relies on the force theorem. 
Therefore, {the} longitudinal spin fluctuations { were neglected}.
{The} details of formulation of the functional integral theory to obtain $\omega(T,{\bf e}_i)$ based on the TB-LMTO method and the KKR methods are summarized in references~\cite{Hiramatsu1,Yamashita2022,Hiramatsu2023,Yamashita2023,Sakuma2024,Yamashita2024-2,Yamashita2025} and in~\cite{StauntonPRL,StauntonPRB}, respectively.

In { this} study, the local spin density approximation (LSDA) was employed with the von~Barth--Hedin--Janak functional as  { the} exchange-correlation potential. 
The lattice constants were set to 0.5875~nm and 0.5905~nm for Mn$_2$VAl~\cite{Weht1999} and Mn$_2$VGa~\cite{Kumar2008}, respectively. 
Spin-orbit coupling in the valence bands was neglected. 
The structure constants were expanded up to $l=2$. The $3d$ electrons of Ga were treated as core states.
The { density of states} (DOS) $D(T,E)$ and the magnetization $M_i(T)$ at site $i$ and at a finite temperature are calculated using $\omega(T,{\bf e}_i)$ { by}
\begin{align}
D(T,E)
&= -\frac{1}{\pi} {\textrm {Im}}
\Bigg[
{\textrm Tr}_{iLs}
\int d{\bf e}_i \,
\omega_i(T,{\bf e}_i)
G(z^+,{\bf e}_i)
\Bigg] ,
\end{align}
\begin{align}
M_i(T)
&= -\frac{1}{\pi} {\textrm {Im}}
\Bigg[
{\textrm Tr}_{Ls}
\int_{-\infty}^{\infty} dE \, f(E,T,\mu) \nonumber \\
&\qquad \times
\int d{\bf e}_i \,
\omega_i(T,{\bf e}_i)
\sigma_z
G_{ii}(z^+,{\bf e}_i)
\Bigg] .
\end{align}
where $G$ is the Green's function of the TB-LMTO method including transverse spin fluctuations. 
$\sigma_z$ is the $z$-component of the Pauli matrix. $f$ is the Fermi-Dirac distribution. 
The trace is taken with respect to the space of the atom indices $i$, orbital indices $L$, and spin index $s$. 
The Kubo-Greenwood formula was used to calculate the conductivities:
\begin{widetext}
\begin{eqnarray}
\label{KG1}
\sigma^{s}_{\nu\nu}(T)
&=& \frac{\hbar}{4\pi V}
\int_{-\infty}^{\infty} dE
\left(-\frac{\partial f(E,T,\mu)}{\partial E}\right) \nonumber \\
&& \times
{\rm Tr}_{iL}
\Big\langle
J_{\nu}
\big(G(z^+,\lbrace{\bf e}\rbrace)-G(z^-,\lbrace{\bf e}\rbrace)\big)
J_{\nu}
\big(G(z^+,\lbrace{\bf e}\rbrace)-G(z^-,\lbrace{\bf e}\rbrace)\big)
\Big\rangle_{\omega(T,{\bf e}_i)} .
\end{eqnarray}
\end{widetext}
Here, $J_{\nu}$ is the current operator in direction $\nu$ and $V$ is the volume of the primitive cell. 
$z^{\pm}$ is defined as $E\pm i \delta$, where $\delta$ { was} chosen { as}  2~mRy in this { study}. 
The parameter $\delta$ acts similarly to impurity scattering independent of spin disorder at finite temperature. 
This value also gives rise to a residual resistivity as in previous works~\cite{Ogura2012,Sakuma2022}. 
$\mu$ { denotes}  the chemical potential of the system. 
The chemical potential { was} determined { such} that the total number of electrons and holes { was} balanced at finite temperatures,
{ was} found by integrating the DOS convoluted by the Fermi--Dirac distribution. 
To calculate Equation~(\ref{KG1}), the vertex correction and coherent terms must be considered.
Several { studies} summarized the details of the treatment of the Kubo--Greenwood formula and vertex correction term in the TB-LMTO method~\cite{Carva2006,Sakuma2022}.
Note that the effect of phonon scattering on the temperature dependence of physical quantities { was not considered in this study}. 
Therefore, the scattering sources for the spin-resolved conductivity { were} attributed only to spin and chemical disorder. 
A $15\times15\times15$ k-points mesh was used for the DLM-CPA calculation to investigate the temperature dependence of magnetization.
For { the} conductivity calculations, $70\times70\times70$ k-points were sampled in the full Brillouin zone. 
A mesh with $30\times30\times30$ k-points was used to calculate  the spin polarization ratio from { the} DOS.

\section{Results and Discussion}

\begin{figure}
\centering
\includegraphics[width=8cm]{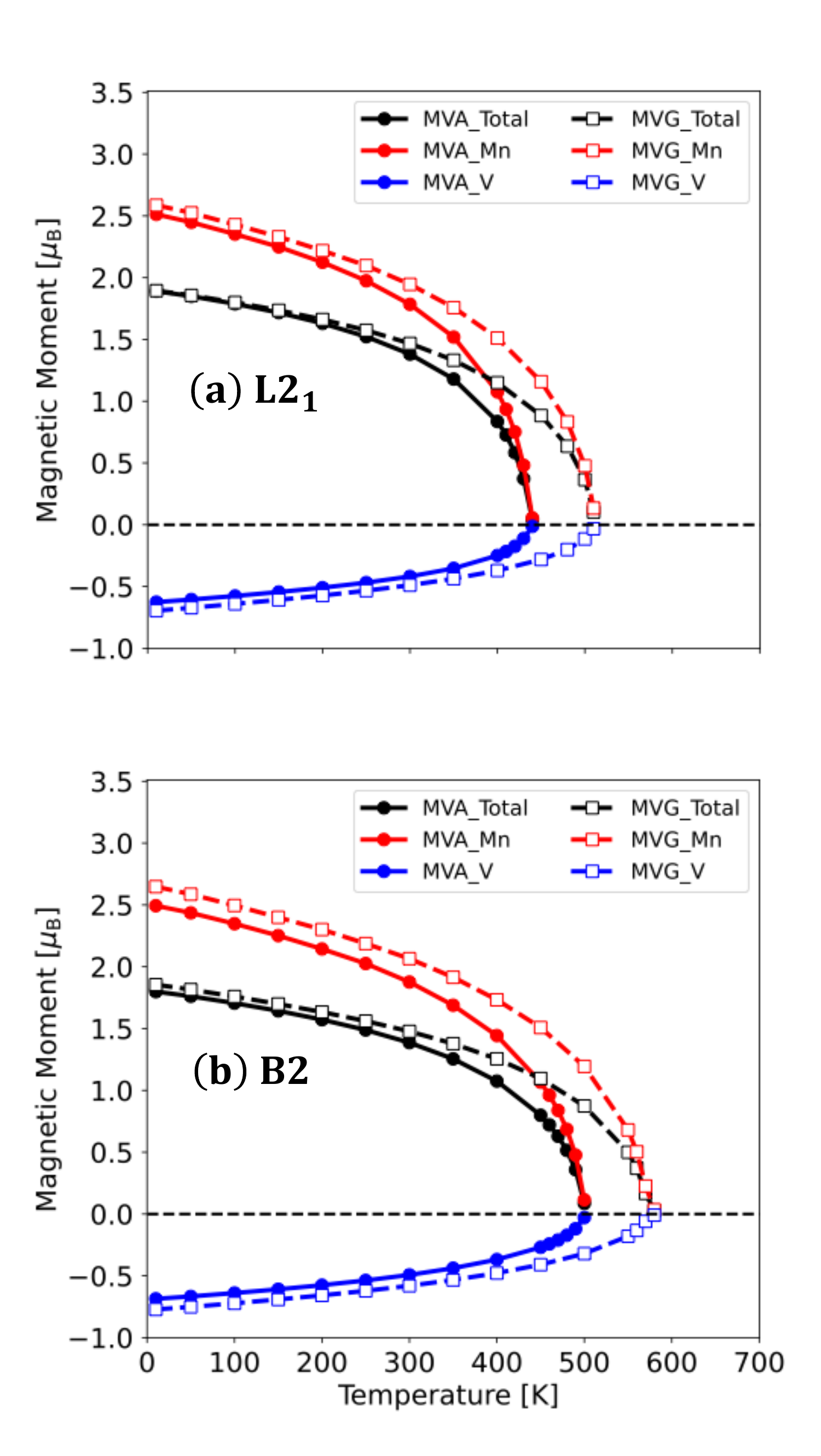}
\caption{Temperature dependence of the total magnetization and site-resolved magnetic moments
         of { the} Mn and V sites of Mn$_2$VAl (MVA, closed circles) and Mn$_2$VGa (MVG, open squares) with (a)
         $L2_1$ and (b) $B2$ structures. Black, red, and blue lines correspond to
         the total magnetization and to Mn and V atoms, respectively. 
         For both structures, the magnetic moments of the Mn atoms are multiplied
         by 2 { because of} the multiplicity of the position.
         Calculations were performed with LSDA.}
\label{MTMVAG}
\end{figure}

Figure~\ref{MTMVAG} shows the temperature dependence of the magnetization of Mn$_2$VAl and Mn$_2$VGa with $L2_1$ or $B2$ structure. 
The calculated magnetization curves resemble a Langevin function { as} a result of the mean-field theory of phase transition, { owing} to the single-site and static approximations that were used.
The calculated Curie temperatures and total magnetic moments at zero temperature are summarized in Table~\ref{DATA}. 
The predicted Curie temperatures { were} 440 and 520~K for Mn$_2$VAl and Mn$_2$VGa with $L2_1$ structure and 500 and 580~K for the alloys with $B2$ structure.
Interestingly, the predicted results indicate that the alloys with $B2$ structure have higher Curie temperatures than { those with the} $L2_1$ structure.

\begin{table}
    \centering
    \caption{Calculated Curie temperatures and total magnetic moments 
		         of Mn$_2$VAl and Mn$_2$VGa at $T=0$.}
    \begin{ruledtabular}
    \begin{tabular}{ccccc}   
        System & $T_{c}^{\rm LDA}$ [K] & $T_{c}^{\rm GGA }$ [K] & $M^{\rm LDA}_{\rm Total}$ [$\mu_B$]  & $M^{\rm GGA}_{\rm Total}$ [$\mu_B$]  \\
    \hline  
     Mn$_2$VAl  $L2_1$  & 440 & 600 & 1.913 & 1.943 \\
     Mn$_2$VGa  $L2_1$  & 520 & 690 & 1.914 & 1.963 \\
     \hline   
     Mn$_2$VAl  $B2$    & 500 & 690 & 1.816 & 1.912 \\
     Mn$_2$VGa  $B2$    & 580 & 790 & 1.872 & 1.986 \\
    \end{tabular}
    \label{DATA}
    \end{ruledtabular}
\end{table} 
Experimentally, {a} Curie temperature of 768 K { was} reported for { the} bulk
Mn$_2$VAl by Umetsu et al.~\cite{Umetsu2015}. For Mn$_2$VGa, Kumar et al. reported a Curie temperature of 783~K for bulk samples~\cite{Kumar2008}. 
Most of our DLM-CPA calculations resulted  in underestimated Curie temperatures compared to the experimental results. 
For comparison, we applied the generalized gradient approximation (GGA) to estimate the Curie temperatures of these systems. 
The following results were obtained: 600 and 690~K for the $L2_1$ structure of Mn$_2$VAl and Mn$_2$VGa, and 690 and 790~K for the B2 structure of Mn$_2$VAl and Mn$_2$VGa, respectively. 
The resulting Curie temperatures {were} still underestimated {in} several cases, although GGA {improved the} theoretical Curie temperatures. 
This {result} may originate from the fact that GGA increases the  energy gap, as discussed by Weht et al. and K\"ubler~\cite{Weht1999,Kubler2006}. 
The present treatment {was} based on the mean-field approximation: thus, it {provides} upper bounds for the Curie temperature~\cite{Kubler2006}. 

Previous theoretical studies attempted to estimate the Curie temperatures of { these} alloys. 
For example, Chico et al.~\cite{Chico2016} calculated the Curie temperature of Mn$_2$VAl and Mn$_2$VGa using an effective classical Heisenberg model combined with the Liechtenstein--Katsnelson--Antropov-Gubanov formula~\cite{Lichtenstein1987} and a Monte Carlo method for several exchange correlation potentials. 
Their results for both LSDA and GGA also underestimated the Curie temperatures as compared with { the} experimental results. 
In another study, Sasioglu et al.~\cite{Sasioglu2005}, have calculated the Curie temperature of Mn$_2$VAl using an effective classical Heisenberg-type spin Hamiltonian. 
They extracted exchange interactions using the frozen magnon approach combined with the mean-field approximation. K\"ubler~\cite{Kubler2006} used a different
approach { for calculating} the Curie temperatures of Mn$_2$VAl using a formula, which is similar to the random phase approximation in the Heisenberg model.
In this case, the exchange interactions were also obtained from {the} spin-spiral excitation energies. 
Both results { also} underestimated the experimental values.

Unlike localized spin Hamiltonian approaches, the present theoretical approach considers the itinerant nature of the $d$ electrons. 
However, the results {were} not significantly different from previous theoretical predictions. 
One possible reason for this {result} is that {proposed} scheme is based on the ASA. 
Additionally, reports have shown that spin-phonon coupling can modify the Curie temperatures of magnetic materials (see, for example~\cite{Tanaka2020,Mankovsky2020}).

\begin{figure*}
\centering 
\includegraphics[width=18cm]{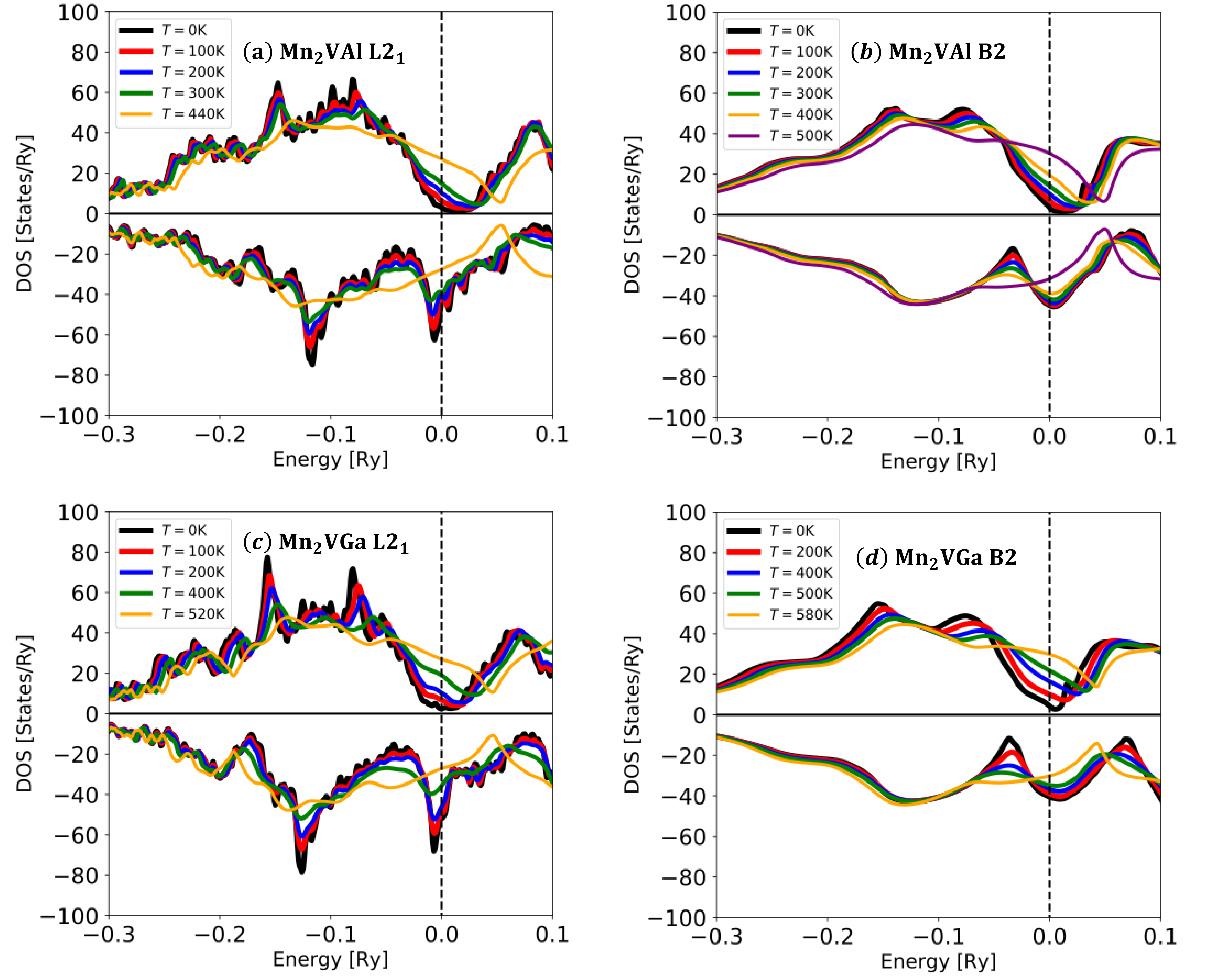}
\caption{Temperature dependence of the DOS of
         (a) $L2_1$- and (b) $B2$- Mn$_2$VAl and (c) $L2_1$- and (d) $B2$- Mn$_2$VGa.
				 Energy zero in { the} horizontal axis was chosen as { the} chemical potential at each temperature.
				  Calculations were performed with LSDA.} 
\label{DOSMVAG}
\end{figure*}

Figure~\ref{DOSMVAG} shows the temperature dependence of  the DOS for Mn$_2$VAl and Mn$_2$VGa in with (a) $L2_1$  and (b) $B2$  structures. 
These figures demonstrate {that} the half-metallic features { of} the DOS deteriorated monotonically with increasing temperature.

Previous theoretical studies have investigated the temperature dependence of the DOS of half-metallic Heusler alloys using { the} dynamical mean-field theory
(DMFT)~\cite{Chioncel2006,Chioncel2008,Chioncel2009} and DLM-CPA~\cite{Lezaic2006,Nawa2020,Kumiawan2021,Yamashita2025}. In most of these
previous studies, the half-metallic character of the DOS { was} destroyed { owing} to the appearance of non-quasi particle states near the Fermi level or the alteration
of the DOS by spin disorder as the temperature increased. 
The same {trend was} observed here for Mn$_2$VAl and Mn$_2$VGa. An exception {was} the case of  Co$_2$MnAl with $L2_1$ structure, {as} predicted by Yamashita et al.~\cite{Yamashita2025}.
Additionally, Chioncel et al. used GGA+DMFT to calculate the temperature dependence of the spin polarization of Mn$_2$VAl from the DOS at the Fermi level~\cite{Chioncel2009}. 
Unlike DLM-CPA, their DMFT approach incorporates the dynamical components of spin fluctuations and non-quasi-particles. 
However, DMFT calculations are not entirely parameter-free {because of a} choice of the Coulomb parameter $U$ is required. 
Nevertheless, their conclusions { were} qualitatively the same as {those of the present study}.

\begin{figure}
\centering 
\includegraphics[width=8cm]{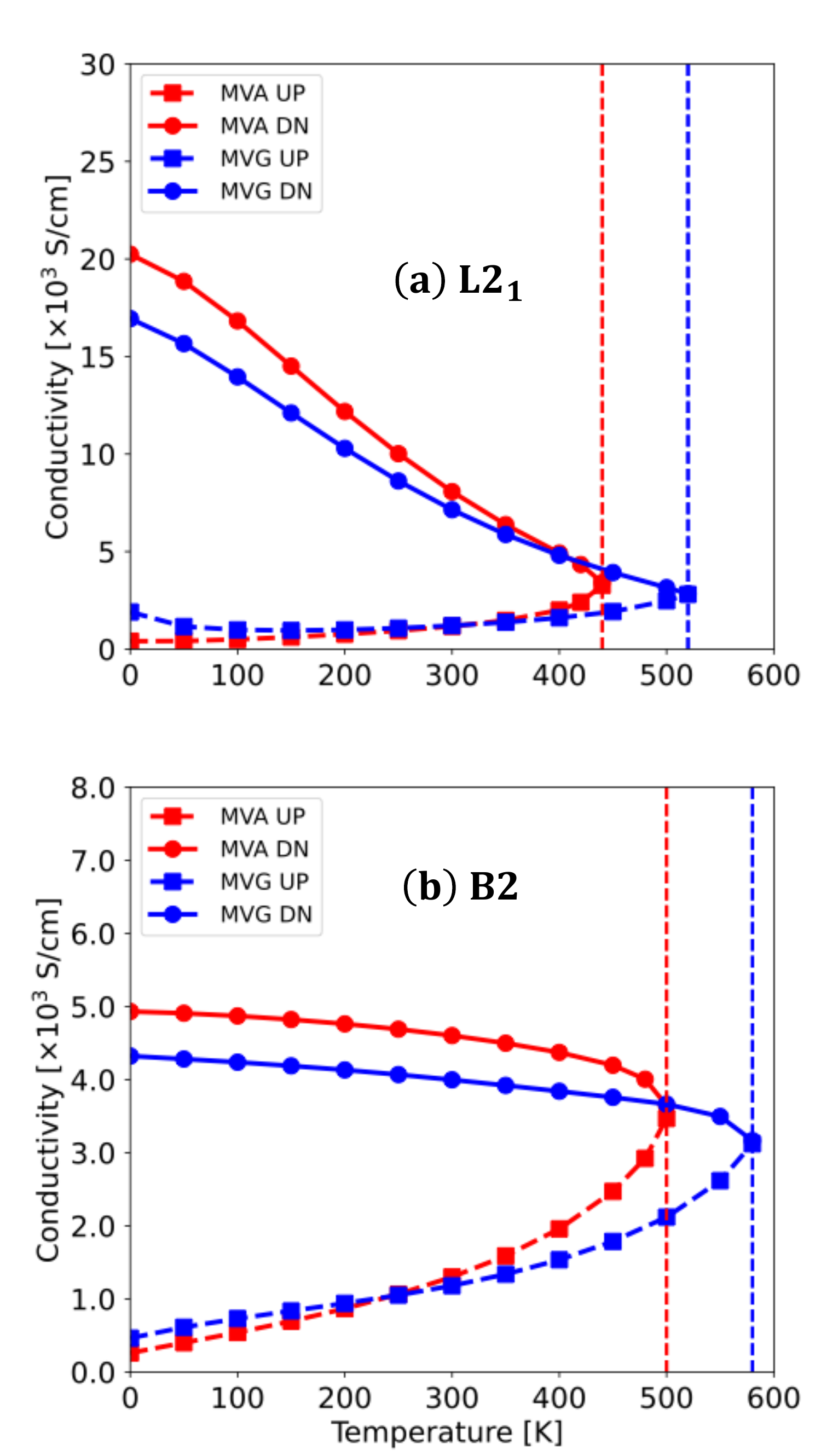}
\caption{Calculated temperature dependence of spin-resolved conductivities of 
         Mn$_2$VAl (red) and Mn$_2$VGa (blue) with the (a) $L2_1$ and (b) $B2$ structures, respectively. 
				 Squares correspond to majority spin channel {UP}  and circles
				 correspond to minority spin channel {DN}. Vertical lines correspond to Curie
         temperatures for each alloy. Calculations were performed with LSDA.}
\label{CMVAG}
\end{figure}

Figure~\ref{CMVAG} shows the temperature dependence of the spin-resolved longitudinal conductivities of Mn$_2$VAl and Mn$_2$VGa { for} both structure types. 
{ As shown in} Figure~\ref{CMVAG} (a), a nearly half-metallic behavior {was} observed close to $T=0$ for both Mn$_2$VAl and Mn$_2$VGa with $L2_1$ structure. 
This { result} is consistent with the calculated DOS at zero temperature.
Unlike the temperature dependence of the DOS, the spin-resolved conductivity of the majority-spin channel of  $L2_1$-Mn$_2$VGa first decreases and then improves at low temperatures. 
This behavior can be explained by the competition between a metallic transition due to an altered DOS and spin-disorder scattering. 
Both phenomena stem from spin transverse fluctuations at finite temperatures. 
The latter may be more significant in the case of $L2_1$-Mn$_2$VGa at low temperatures and could lead to an effective improvement in spin polarization. 
States that appear near the chemical potential at finite temperatures are composed of $d$ states, which have lower mobility than $s$ or $p$ states. 
This trend in Mn$_2$VGa can also be { observed} when $\delta$ is chosen to be 1~mRy.

These results indicate that the DOS and conductivity behaviors at finite temperatures are inconsistent with each other with respect to half-metallicity.
This finding is consistent with {those in} several previous studies~\cite{Soulen1998,Dowben2011,GHFbook}. 
This {result} can be understood from the fact that {the} conductivities depend on both the DOS and the relaxation times or velocity operator. 
For the $B2$ structure, as shown in Figure~\ref{CMVAG} (b), the obtained conductivities in the majority spin channel increased as expected, {according to} the temperature dependence of the DOS. 
For both alloys and with both structures, the conductivities of the minority spin channel behaved {similar to} those of typical magnetic metals. 
{ The} contributions from {the} vertex correction term { were} negligible for the  $L2_1$ structure. 
However, for the $B2$ structure, {the} vertex correction term significantly { affected} the conductivity.

Once the temperature dependence of the longitudinal spin-resolved conductivity is determined, the temperature dependence of the spin polarization 
$P_{\sigma}(T)$ can be defined as follows:

\begin{equation}
P_{\sigma}(T)=\frac{\sigma^{\rm up}_{\nu\nu}(T)-\sigma^{\rm dn}_{\nu\nu}(T)}
              {\sigma^{\rm up}_{\nu\nu}(T)+\sigma^{\rm dn}_{\nu\nu}(T)} \times100.
\end{equation}

\begin{figure*}
\centering 
\includegraphics[width=18cm]{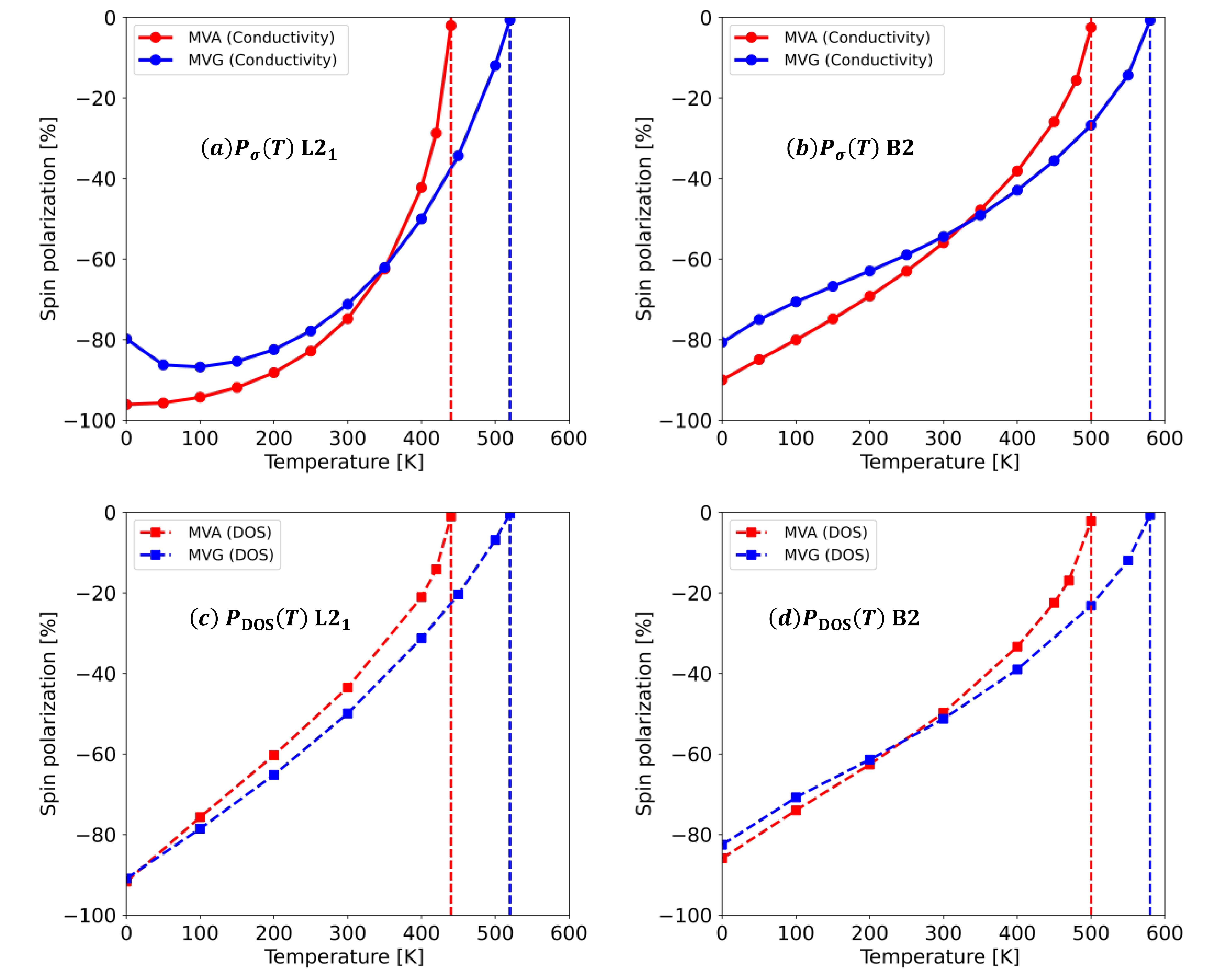}
\caption{Calculated temperature dependence of spin polarization calculated from (a),  (b)  spin resolved conductivities
         and (c), (d) DOS of Mn$_2$VAl (red) and Mn$_2$VGa (blue) {with the}  $L2_1$ and $B2$ structures, respectively. 
				 Vertical lines correspond to Curie temperatures for each alloy. Calculations were performed with LSDA. }
\label{PMVAG}
\end{figure*}

The corresponding results are {presented} in Figure~\ref{PMVAG}. {It} also shows the temperature dependence of the spin polarization $P_{\rm DOS}(T)$, which { was} evaluated from the DOS at the chemical potential, for comparison.  
The equation for spin polarization from the DOS is defined as

\begin{equation}
  P_{\textrm{DOS}}(T)=\frac{D^{\rm up}(T,\mu)-D^{\rm dn}(T,\mu)}
	                         {D^{\rm up}(T,\mu)+D^{\rm dn}(T,\mu)}\times100.
\end{equation}

Here, $\delta$ was set to 1~mRy to calculate the DOS.
Reflecting the results of the temperature dependence of { the} spin-resolved conductivities, the spin polarization ratio $P_{\sigma}(T)$ of Mn$_2$VGa with { the} $L2_1$ structure
improves at low temperatures. 
Unlike $P_{\sigma}(T)$, $P_{\rm DOS}(T)$ of Mn$_2$VGa decreases monotonically with increasing temperature. 
{However, for the} $B2$ structure, the spin polarization ratio decreased monotonically with increasing temperature when evaluated from both {the} conductivity {and}  DOS.
{In addition}, the difference in the spin polarization when evaluated from {the} conductivity or DOS {was} small for the $B2$ structure.

It should be emphasized that the temperature dependence of the spin polarization, as evaluated from {the} temperature dependence of the spin-resolved
conductivity of Mn$_2$VAl with $L2_1$ structure, {was} more robust than that { with} the $B2$ structure. 
Furthermore, the difference in temperature dependence between these two spin polarizations may affect the temperature dependence of { the} giant and tunneling magnetoresistance differently.

It is {noteworthy} that the effects of phonon excitation {and} spin disorder of conductivity and spin polarization can be considered within the CPA framework\cite{Ebert2011,Kodderitzsch2013,Ebert2015,Kou2018,Shinya2020,Wagenknech2017-1,Wagenknech2017-2}. 
However, this { aspect} is beyond the scope of the present {study}.

\begin{figure}
\centering
\includegraphics[width=8cm]{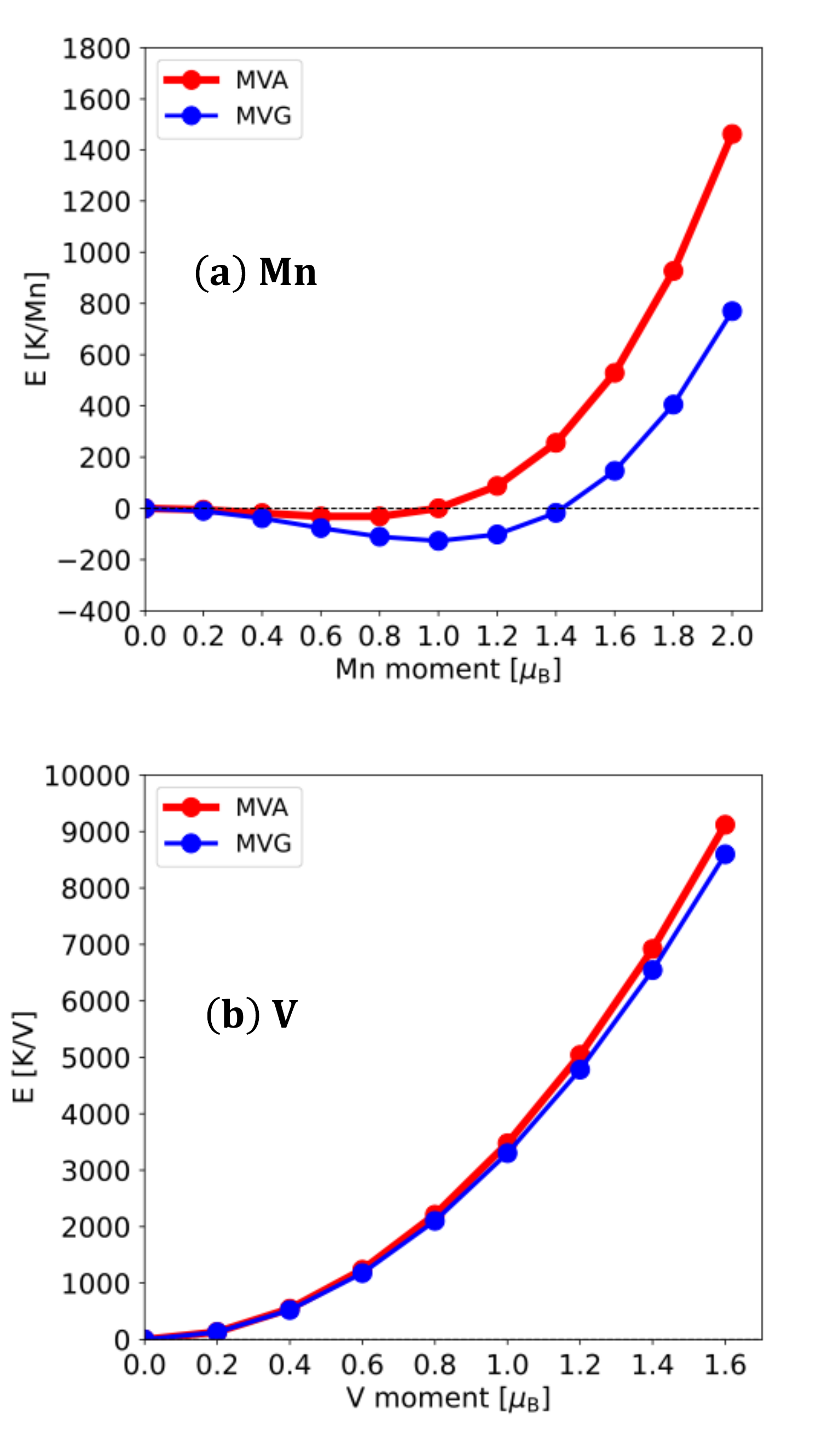}
\caption{Calculated total energy of paramagnetic states with different amplitude of magnetic moment of 
         Mn$_2$VAl (red) and Mn$_2$VGa (blue) {with the} $L2_1$ structures for (a) Mn and (b) V sites,
				 respectively. Calculations were performed with LSDA.}
\label{LSFMVAG}
\end{figure}

Finally, the details of the energy landscape were investigated to examine whether the DLM-CPA scheme based on the force theorem is appropriate for these
alloys. The total energies of the paramagnetic states {were} calculated self-consistently. 
Only the up- and down- spin states were included in the CPA calculations.  
The calculations were performed for Mn$_2$VAl and Mn$_2$VGa with { the} $L2_1$ structure.
In these total energy calculations, the size of the magnetic moment of one magnetic atom {was} varied arbitrarily using the fixed-spin-moment method while allowing the magnetic moments of the other atoms to converge freely~\cite{Khmelevskyi2022}. 
The results are { presented} in Figure~\ref{LSFMVAG}.

Mn atoms have larger magnetic moments at zero temperature. 
However, the total energy landscape {was} shallow, as shown in Figure~\ref{LSFMVAG}(a). 
The results for the V atoms are shown in Figure~\ref{LSFMVAG}(b). 
The magnetic moment of V completely vanished in the paramagnetic state for both structures.

In addition to the LSDA method, the GGA method was used to calculate the total energy landscape of the alloys in the paramagnetic state. 
The results obtained {were} qualitatively similar to those {obtained} from { the} LSDA.
These results suggest that the {atoms in the alloys} may exhibit significant longitudinal spin fluctuations.
For Mn, the thermal fluctuations of the magnetic moments {were} {approximately} 700 K, which is close to the experimental Curie temperature. 
This { property} could lead to a wide range of size fluctuations { in} the magnetic moments. 
Therefore, the magnetic moments may differ from those corresponding to the energy minimum of the total energy landscape. 
Conversely, for V, the magnetic moments are induced and sustained by thermal fluctuations \cite{Hasegawa3, Heine1981, Khmelevskyi2018}. 
However, the results suggest that the magnetic moments of V {were} more itinerant than those of Mn. 
Nevertheless, X-ray spectroscopy experiments have shown an opposite trend.

Several authors have {reported} that incorporating longitudinal fluctuations can change the Curie temperatures of
materials~\cite{Ruban2007,Khmelevskyi2016,Delceg2023}. 
For example, Khemelevskyi~\cite{Khmelevskyi2016} {reported} that longitudinal spin fluctuations can increase the critical temperature. 
The critical temperature of V$_3$Al was underestimated using a Heisenberg-type model that considered only transverse spin fluctuations. 
Antiferromagnetic V$_3$Al is isostructural {with} Mn$_2$VAl and Mn$_2$VGa. 
The present functional integral approach, based on the force theorem, roughly corresponds to a saddle-point approximation that fixes the size of the magnetic moments at the ground-state values. 
However, total energy landscape calculations suggest that the most stable size of magnetic moments differs between the paramagnetic and ground states. 
The results also suggest the existence of a wide range of thermal fluctuations with respect to the size of the magnetic moments.
These results imply that the force theorem approach may not be suitable for alloys, particularly near the Curie temperature. 
A treatment that reflects the energy landscape with respect to { the} magnetic moment amplitude {can} improve {the} theoretical results. 
This {approach} could be {particularly} important for predicting the Curie temperature and describing behavior near it.

\section{Summary and conclusion.}

In { this study}, the temperature dependence of the magnetic properties,
and electronic structure, as well as its effects on spin-resolved transport properties, of Mn$_2$VAl and Mn$_2$VGa { were} investigated. 
Calculations were performed for ordered alloys with  $L2_1$ structure and disordered alloys with  $B2$ structure. 
The DLM-CPA approach used here to express the transverse spin fluctuations  can also  incorporate both the itinerant and localized properties of the $d$ electrons.

The predicted Curie temperatures { were} lower than the experimental values in most cases, which is consistent with the findings of other { studies}. 
Additionally, it was confirmed that the calculated Curie temperatures of { the} alloys with a $B2$ structure { were} higher than the of compounds with an $L2_1$ structure.

The calculated temperature dependence of the DOS shows a monotonic decrease in spin polarization with increasing temperature. 
However, in contrast to the prediction based on the behavior of the DOS at finite temperatures, the temperature dependence of the spin polarization calculated from the conductivity of Mn$_2$VGa with the $L2_1$ structure resulted in a minimum at approximately 100~K. 
This { result} may be considered a transient recovery of half-metallicity at that temperature, and the minimum can be explained by the competition between { the} {metallic transition} due to the alteration of the DOS and spin scattering. 
Both originate from spin disorder at finite temperatures.

The conductivity results for the $L2_1$ structure also indicate that the DOS and conductivity behavior are not necessarily consistent with each other at finite
temperatures. { However}, the temperature dependence of the spin polarization in the $B2$ structures was nearly the same, showing typical half-metallic behavior in { terms of} conductivity and DOS.

Finally, the total energy landscape in { the} paramagnetic states was investigated to examine the longitudinal stiffness of { the} magnetic moments at different sites. The
calculations were performed using the fixed-spin moment method. These results indicate that the materials exhibit strong longitudinal spin fluctuations, suggesting { the} itinerant nature of the $d$ electrons. This { behavior} was also observed for both Mn and V magnetic ions, demonstrating that the magnetic properties of Heusler compounds and alloys cannot be simply classified as localized or itinerant.

\bigskip
\begin{acknowledgments}

This  { study} was financially supported by the European Commission (EC)
European Research Council (ERC) Advanced Grant {\it 'Strain-Free All Heusler
Alloy Junctions--SAHAJ'} (No. 101097475). 
S.Y. acknowledges Dr. Ryoya Hiramatsu, Prof. Dr. Akimasa Sakuma,
and Dr. Sergii Khmelevskyi for valuable discussions, and Prof. Dr. Andrei Ruban
for providing the KKR-ASA program. 
A.H. is grateful for support from the Japan Science and Technology Agency (JST)
Adopting Sustainable Partnerships for Innovative Research Ecosystem (ASPIRE)
program (No. JPMJAP2409) and the Ministry of Education, Culture, Sports, Science
and Technology (MEXT) X-nics (Grant No. JPJ011438).

\end{acknowledgments}

\section*{Data Availability Statement}

Data supporting the findings of this study are available within the
article. If data are not included, they cannot be made publicly available. 
Because no suitable repository exists for hosting data in this field of study.
Additional data are available from the corresponding author upon
reasonable request.


\newpage
\bibliographystyle{unsrt}

\end{document}